\documentclass[10pt,twocolumn,superscriptaddress,aps,prl,balancelastpage]{revtex4-1}
\usepackage{amsfonts,amsmath,amssymb,graphicx,epstopdf,verbatim,dsfont,color}
\usepackage[english]{babel}
\usepackage{subfigure}
\usepackage{epstopdf}
\usepackage{hyperref}

 \graphicspath{{figs/}}
 
\begin{document}

\title{Entanglement entropy scaling transition under competing monitoring protocols}
\author{Mathias Van Regemortel}
\email{mvanrege@umd.edu}

\author{Ze-Pei Cian}

\author{Alireza Seif}

\author{Hossein Dehghani}

\author{Mohammad Hafezi}
\affiliation{
Joint Quantum Institute, College Park, 20742 MD, USA}
\affiliation{The Institute for Research in Electronics and Applied Physics,
University of Maryland, College Park, 20742 MD, USA}
\date{\today}

\begin{abstract}
Dissipation generally leads to the decoherence of a quantum state. In contrast, numerous recent proposals have illustrated that dissipation can also be tailored to stabilize many-body entangled quantum states. While the focus of these works has been primarily on engineering the non-equilibrium steady state, we investigate the build-up of entanglement in the quantum trajectories. Specifically, we analyze the competition between two different dissipation channels arising from two incompatible continuous monitoring protocols. The first protocol locks the phase of neighboring sites upon registering a quantum jump, thereby generating a long-range entanglement through the system, while the second destroys the coherence via a dephasing mechanism. By studying the unraveling of stochastic quantum trajectories associated with the continuous monitoring protocols, we present a transition for the scaling of the averaged trajectory entanglement entropies, from critical scaling to area-law behavior. Our work provides an alternative perspective on the measurement-induced phase transition: the measurement can be viewed as monitoring and registering quantum jumps, offering an intriguing extension of these phase transitions through the long-established realm of quantum optics. 
\end{abstract}

\maketitle

 While coupling a quantum system with the environment is often detrimental for preserving entanglement \cite{gardiner2004quantum}, dissipation can also be engineered and utilized to stabilize exotic and highly entangled many-body states \cite{diehl2008quantum,kraus2008preparation,verstraete2009quantum}. With the development of recent experimental platform, such as circuit quantum electrodynamics (QED) \cite{blais2007quantum,schoelkopf2008wiring,houck2012chip,kounalakis2018tuneable} and Rydberg polaritons \cite{saffman2010quantum}, strongly entangled photonic states can be engineered with reservoir engineering \cite{poyatos1996quantum} and tailoring dissipation schemes \cite{weimer2010rydberg,kastoryano2011dissipative,cho2011optical,barreiro2011open,reiter2013steady,cian2019photon}.

Quantum phase transitions \cite{sachdev2007quantum} typically come with different phases for entanglement entropy, as shown for the exemplary Bose-Hubbard model \cite{lauchli2008spreading}, where numerous works investigated the scaling of correlations with the system size \cite{fisher1989boson,kuhner1998phases,ejima2011dynamic}.
Also local projective measurements of a quantum state destroy the entanglement generated by unitary evolution, which may lead to a phase transition of entanglement entropy across the system. A number of recent works have explored quantum circuits of random unitaries alternated with local measurements, and a phase transition was seen for the scaling of entanglement entropy \cite{nahum2017quantum,li2018quantum,skinner2019measurement,chan2019unitary,Vasseur2019,choi2020quantum,gullans2020scalable,jian2020measurement,zabalo2020critical,bao2020theory,shtanko2020classical}. Later a similar transition was reported for the stochastic trajectories from quantum systems under a local continuous-monitoring protocol, which induces an interplay with the entanglement from the unitary dynamics of the Hamiltonian \cite{fuji2020measurement,cao2019entanglement,alberton2020trajectory}. More generally, it is worth investigating whether stochastic quantum trajectories, a well-established quantum optics formalism \cite{dalibard1992wave,wiseman2012dynamical}, can provide more insight into measurement-induced phase transitions, by employing the possibility of registering the quantum jumps.

\begin{figure}
    \centering
    \includegraphics[width=\columnwidth]{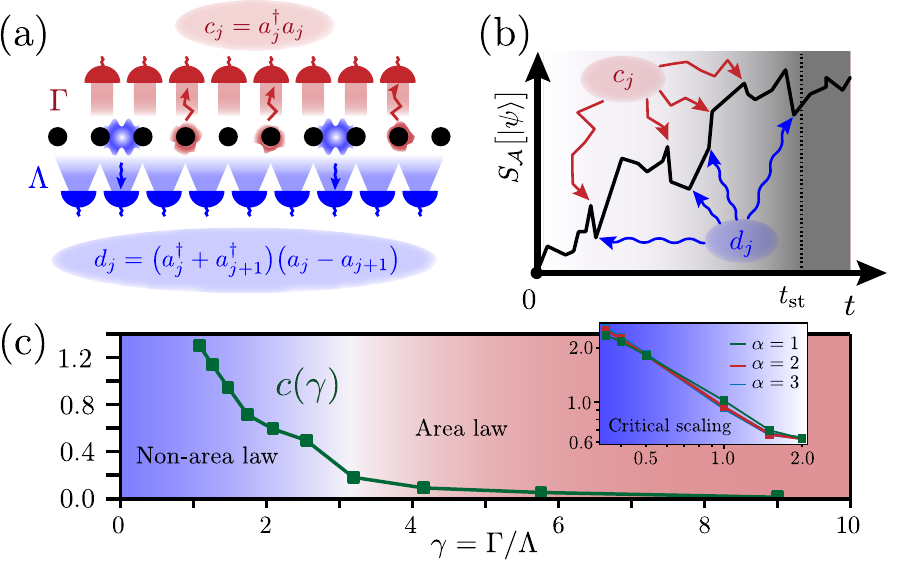}
    \caption{A schematic illustration of our setup and the scaling of trajectory entanglement entropy. (a) We analyze the stochastic evolution of the system under continuous monitoring with two competing monitoring protocols, characterized by the registering of jump operators $d_j$ and $c_j$ with rates $\Lambda$ and $\Gamma$, respectively. (b) The quantum state $|\psi(t)\rangle$, starting from zero entropy, follows a stochastic trajectory under the continuous monitoring with $d_j$ and $c_j$, which can be seen as random fluctuations of entanglement entropy of a subsystem. Over long enough times $t_\text{st}$, the system is expected to converge to a steady state. (c) The fitting parameter from Eq. \eqref{eq:S_cft} $c(\gamma)$, with $\gamma \equiv \Gamma/\Lambda$, obtained from fitting to a system with $L=32$, showing a transition from area law (high $\gamma$) to non-area law (low $\gamma$). The inset shows $c(\gamma)$, derived from $c_\alpha(\gamma)$ from the Renyi entropy of order $\alpha$ for CFT's \eqref{eq:c_alpha}. The coincidence of the $c(\gamma)$ curves for different order $\alpha$ for small $\gamma$ is suggestive for a phase of critical scaling, where the effective central charge shows the onset of a power-law divergence as function of $\gamma$. }
    \label{fig:Fig1}
\end{figure}

Here we present a scaling transition of entanglement entropy in a quantum system, governed entirely by dissipative dynamics -- coming from the interplay of two continuous monitoring protocols -- in the \emph{absence} of unitary dynamics. In Fig. \ref{fig:Fig1}(a), we illustrate the model; a chain of bosonic modes, of length $L$ and with open boundaries, is first monitored with a protocol that locks the phase of two adjacent sites with jump operators
\begin{equation}
\label{eq:ch}
d_{j} \equiv \big(a^\dagger_j + a^\dagger_{j+1} \big) \big( a_j - a_{j+1} \big),
\end{equation}
where $a_j$ ($a^\dagger_j$) is the annihilation (creation) operator for the bosonic mode on site $j$ \cite{diehl2008quantum}. The second monitoring protocol is dephasing, with jump operators 
\begin{equation}
\label{eq:cd}
c_j \equiv a^\dagger_{j}a_{j}.
\end{equation}
 The rates of the monitoring for phase-locking $d_j$ and dephasing $c_j$ are given by $\Lambda$ and  $\Gamma$, respectively. We investigate the competition between the two monitoring schemes in terms of the \emph{reduced} dephasing rate
\begin{equation}
\label{eq:gamma}
    \gamma \equiv \frac{\Gamma}{\Lambda}.
\end{equation}

The continuous monitoring and the recording of the jumps is a crucial element of this work. While dissipation is often introduced to account for the decoherence of a quantum state, we elaborate specific implementation schemes that allow for the continuous tracking of the system in a circuit QED setup \cite{supp}. The random occurrence and detection of the quantum jumps \eqref{eq:ch} and \eqref{eq:cd} implies that the dynamics of a quantum state $|\psi(t)\rangle$ is inherently stochastic, as depicted in Fig. \ref{fig:Fig1}(b). To characterize the state of the system $|\psi(t)\rangle$, we use the entanglement entropy of a subsystem $\mathcal{A}$, a \emph{state-dependent} quantity, which is evaluated as $S_\mathcal{A} [|\psi(t)\rangle]= -\text{Tr} \rho_\mathcal{A} \log \rho_\mathcal{A}$ with $\rho_\mathcal{A}$ the reduced density matrix of the state $|\psi(t)\rangle$ on $\mathcal{A}$. It is crucial that $S_\mathcal{A}$ is a strongly \emph{nonlinear} function of the stochastic states $|\psi(t)\rangle$. As an immediate consequence, statistical averages of $S_\mathcal{A} [|\psi(t)\rangle]$ over the states can \emph{not} be retrieved from a master-equation approach. This in stark contrast with linear quantities, such as operator expectation values $\langle O \rangle_t = \langle \psi(t)| O | \psi(t)\rangle$ \cite{dum1992monte,dalibard1992wave}, which converge to the master equation and exhibit a notion of ergodicity \cite{kuemmerer2004pathwise} and thermalization \cite{schachenmayer2014spontaneous,ashida2018thermalization}. Importantly, there is a convergence time $t_\text{st}$ for $S_\mathcal{A}$, after which the stochastic state $|\psi(t)\rangle$ is sampled from a steady-state distribution.

We present a scaling transition for the averaged entanglement entropy of stochastic states, after some evolution time, across a critical value of the reduced dephasing rate \eqref{eq:gamma}, as presented in Fig. \ref{fig:Fig1}(c). When phase-locking dominates, the state is superfluid and entanglement entropy has a strong dependence on subsystem size. We report the critical scaling, characterized by an effective central charge $c(\gamma)$ (green line). While an appropriate scaling analysis is difficult due to numerical constraints, our hypothesis is motivated by the results on the inset of Fig. \ref{fig:Fig1}c: the effective curvature $c_\alpha$ found for averaged higher-order R\'enyi entropies $S_\alpha = 1/(1-\alpha) \log{ \text{Tr} \rho^\alpha}$ follows the universal scaling from conformal field theory (CFT) \cite{calabrese2004entanglement,fagotti2011entanglement}
\begin{equation}
\label{eq:c_alpha}
    c_\alpha(\gamma) = \frac{c(\gamma)}{2}\Big(1+\frac{1}{\alpha}\Big),
\end{equation}  
and the effective central charge $c(\gamma)$ shows the onset of a power-law divergence for $\gamma\rightarrow 0$. While the \emph{full} master equation in this regime is expected to converge to a mixed steady state with a volume law, there is no \emph{a priori} reason why \emph{trajectory} entanglement entropy should follow the same scaling.

When dephasing becomes more important (higher $\gamma$), the scaling changes to an area law, marked by $c(\gamma)\approx0$. Intuitively, the transition can be further understood from an order parameter in a simple Gutzwiller picture, elaborated in \cite{supp}. In circuit-models, two incompatible types of measurements \emph{without} unitary entangling gates can also lead to a scaling transition for entanglement entropy \cite{lavasani2020measurement,ippoliti2020entanglement}. Our work aims to extend the recent understanding of a measurement-induced phase transition, as seen in discrete random circuits, to the stochastic trajectories of an unraveling associated with the continuous monitoring of a quantum system.

\textit{Stochastic trajectories.--} The system dynamics is fully governed by the two competing monitoring protocols. A state $|\psi(t)\rangle$ then follows a stochastic trajectory, as was originally introduced in the seminal works \cite{dalibard1992wave,dum1992monte} as a way to sample the master equation of an open quantum system. Whereas the \emph{unraveling} for sampling a master equation is not unique, here it relates unequivocally to the monitoring protocol presented in Fig. \ref{fig:Fig1}(a)-(b), thereby relying explicitly on the hypothesis of detector-dependent stochastic pure-state dynamics \cite{wiseman2012dynamical}.

The sampling of quantum trajectories from the continuous monitoring goes as follows. At time $t$, we evaluate whether there is a jump in the differential time interval $[t,t+\Delta t]$ by evaluating the probability $\Delta p = \sum_{j=1}^{L-1} \Delta p^{(d)}_j + \sum_{j=1}^{L} \Delta p^{(c)}_j$, a summation over the probabilities $\Delta p^{(d)}_j$ and $\Delta p^{(c)}_j$ of the jumps $d_j$ and $c_j$ to occur, with $\Delta p^{(b)} = \gamma^{(b)} \langle \psi(t) | b^\dagger b |\psi(t) \rangle \Delta t$ and $\gamma^{(b)} = \{\Lambda,\Gamma\}$, accordingly.

If no jump is detected (probability $1-\Delta p$) we evolve the state over $\Delta t$ with the anti-Hermitian Hamiltonian $H_\text{eff} = - \frac{i \Lambda }{2}\sum_{j=1}^{L-1} d^\dagger_j d_j - \frac{i \Gamma }{2} \sum_{j=1}^{L} c^\dagger_j c_j $. If a jump is recorded (probability $\Delta p$), we select one $b\in \{d_j,c_j\}$ with probabilities $\Delta p^{(d)}_j$ or $\Delta p^{(c)}_j$, respectively, to evaluate $|\psi(t+\Delta t)\rangle = b|\psi(t)\rangle$.

After each time step $\Delta t$, the state $|\psi(t) \rangle$ is normalized to simulate the stochastic evolution of $|\psi_i(t)\rangle$ in the monitoring scheme. Importantly, both detection (probability $\Delta p$) as well as \emph{absence} of a jump (probability $1-\Delta p$) in $\Delta t$ yields information about the state of the system to an observer. This was illustrated in several recent experiments to monitor the stochastic evolution of a superconducting qubit \cite{murch2013observing,weber2014mapping,sun2014tracking,minev2019catch} and how simultaneously monitoring dephasing and relaxation leads to an interplay \cite{ficheux2018dynamics}.

The phase-locking \eqref{eq:ch} stabilizes a pure Bose-Einstein Condensate (BEC) dark state with long-range entanglement, where all $L$ particles are injected in the zero-momentum mode; $|D\rangle = (a^{\dagger}_{k=0})^L|0\rangle$, with $a^{\dagger}_k$ the creation operator of a photon with momentum $k$ \cite{diehl2008quantum,kraus2008preparation}, while dephasing \eqref{eq:cd} directs the system to a product of local Fock states with zero entanglement.

While the local $U(1)$ symmetry is broken by the phase locking \eqref{eq:ch}, a global $U(1)$ symmetry is present in our system; both jumps $d_j$ (\ref{eq:ch}) and $c_j$ (\ref{eq:cd}) conserve the total particle number. For the upcoming analysis we fix the filling factor $n=1$ and the evolution starts from the Fock state $|\psi(t=0)\rangle = |...1111...\rangle$.

\textit{Gutzwiller approach.--} Given a stochastic trajectory state $|\psi(t)\rangle$, upon taking the thermodynamic limit $L \rightarrow \infty$, we can study the dynamics of on-site observables in the Gutzwiller approximation by considering a mean-field coupling to neighboring sites for the single-site reduced density matrix \cite{lebreuilly2019stabilizing,casteels2017optically}. An effective single-site Liouvillian can be constructed for the Gutzwiller master equation of the reduced density matrix after averaging over trajectories. A full numerical analysis of the mean-field \emph{order parameter}  $\alpha \equiv \langle a \rangle $ shows that it vanishes across a critical value $\gamma_c^\text{(GW)} \approx 3 $, as such providing a suggestive sign for a trajectory transition, see \cite{supp}.


\textit{Trajectory entanglement entropy.--} We focus on evaluating the Von Neumann entanglement entropy of the trajectory states from a system of size $L$, $| \psi_L \rangle$, in a subsystem $\mathcal{A}$ containing $l$ sites from the left: $S(l)[| \psi_L \rangle] = - \text{Tr} \big[ \rho_\mathcal{A} \log \rho_\mathcal{A} \big]$, with $\rho_\mathcal{A} = \text{Tr}_{\mathcal{B}} |\psi_L\rangle \langle \psi_L |$ the reduced density matrix of $\mathcal{A}$ and $\mathcal{B}$ containing the remaining $L-l$ sites. We evaluate the averaged entanglement entropy of a set of $M$ stochastic trajectory states $| \psi^{(\gamma)}_L(t) \rangle_i$, $i\in [1,M]$, at time $t$ in a system with reduced dephasing rate $\gamma$ \eqref{eq:gamma},
\begin{equation}
\label{eq:S_av}
\overline{S}^{(\gamma)}_L(l,t) = \frac{1}{M}\sum_{i=1}^{M} S(l)[| \psi^{(\gamma)}_L(t) \rangle_i].
\end{equation}

\begin{figure}
    \centering
    \includegraphics[width=\columnwidth]{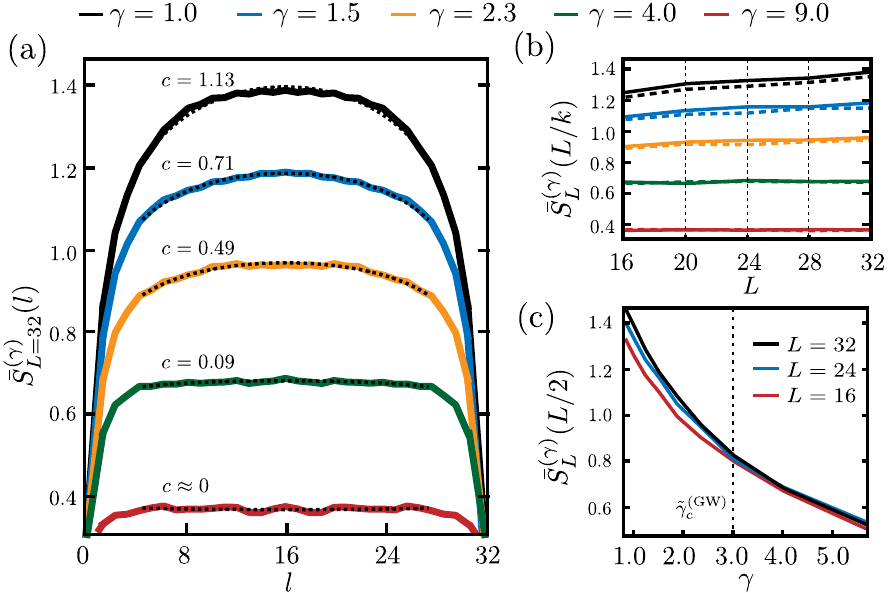}
    \caption{Different scalings of $\bar{S}^{(\gamma)}_L(l)$ from averaging over $10000$ steady-state trajectory states. (a) The scaling of $\bar{S}^{(\gamma)}_{L=32}(l)$ as function of $l$ for different values of the reduced dephasing $\gamma$. A transition is seen from critical scaling (black, blue, orange lines) to an area law (green and red line), as obtained from the effective central charges $c(\gamma)$ found by fitting (dotted lines) the functional form \eqref{eq:S_cft} (b) The scaling of $\bar{S}^{(\gamma)}_{L}(L/2)$ (solid) and $\bar{S}^{(\gamma)}_{L}(L/4)$ (dashed) as function of $L$. (c) The dependence of $\bar{S}^{(\gamma)}_{L}(L/2)$ on $\gamma$ for different system sizes $L$; we distinguish a critical point $\gamma_c$ where the lines start to coincide, close to the Gutzwiller critical point $\gamma_c\approx 3$.}
    \label{fig:Fig2}
\end{figure}

\textit{Numerical results.--} We use Matrix Product States (MPS) \cite{perez2007matrix} to sample the stochastic quantum states \cite{daley2014quantum} with the C++ package \emph{ITensor} \cite{ITensor}.

In Fig. \ref{fig:Fig2}, the scaling of the averaged entanglement entropy $\bar{S}_L^{(\gamma)}(l)$ for trajectories sampled from the steady state is illustrated for three parameters $l$ (a),  $L$ (b) and $\gamma$ (c). In Fig. \ref{fig:Fig2}(a) we see that the curves $\bar{S}_L^{(\gamma)}(l)$ show a transition from a strong concave behavior as function of $l$ when phase-locking dominates (black, blue and orange line) to a regime with an area-law behavior (green and red line). After numerical analysis, we identify the scaling of the curves in the phase-locking regime as logarithmic, reminiscent of the scaling of entanglement entropy for ground states of critical Hamiltonians with open boundary conditions \cite{lauchli2008spreading}, given by a result from CFT \cite{calabrese2004entanglement},
\begin{equation}
\label{eq:S_cft}
\bar{S}^{(\gamma)}_{L}(l) = \frac{c(\gamma)}{6} \log \Big[ \frac{2L}{\pi} \sin\Big(\frac{\pi l}{L}\Big)\Big] + s_{0}(\gamma).
\end{equation} 
Here $c(\gamma)$ is the effective central charge and $s_0(\gamma)$ the residual entropy.

A fitting procedure (dotted lines) with the functional form \eqref{eq:S_cft} gives the parameters $c(\gamma)$ (indicated above the curve) and $s_0(\gamma)$, in close agreement with the numerical results (solid lines). In Fig. \ref{fig:Fig1}(c) we summarize our key result: the fitted central charge shows a transition from a nonzero value to zero upon increasing the effective dephasing rate $\gamma$. Consequently, we report a transition from critical scaling of entanglement \eqref{eq:S_cft} to an area law, characterized by $c(\gamma)=0$, which has a plateau value $s_0(\gamma)$ for the bulk entanglement entropy. In the limit $\gamma\rightarrow \infty$ no entanglement can build up and $\bar{S}^{(\gamma)}_L(l) \rightarrow 0$, so that also $s_0(\gamma)\rightarrow 0$.

In the inset of \ref{fig:Fig1}(c), we analyze the critical behavior more closely by investigating R\'enyi entropies of order $\alpha$, which satisfy the universal relation for CFT \eqref{eq:c_alpha} \cite{calabrese2004entanglement, fagotti2011entanglement}. The central charge $c(\gamma)$ is shown, as obtained from $c_\alpha(\gamma)$ of the R\'enyi entropy $S_\alpha$, averaged over steady-state trajectories, analogous to \eqref{eq:S_av}. We conclude that the central charges $c(\gamma)$ coincide within numerical precision, as such retrieving the universality relation from CFT and confirming the reported critical scaling. Moreover, as we let $\gamma \rightarrow 0$, the central charge $c_1(\gamma)$ appears to show a power-law divergence, as was also seen for free fermion trajectories with dephasing \cite{alberton2020trajectory}. 

The exact critical $\gamma_c$ for the scaling transition is difficult to extract from our numerical data. We are computationally limited (mainly the finite bond dimension and local Fock-space truncation of the MPS) to sampling system sizes of $L\lesssim 32$ with $\gamma \gtrsim 0.35$, making a finite-size scaling analysis difficult. We also leave it as an open question whether the power-law scaling for $c(\gamma)$ persists or stabilizes to a finite value at $\gamma > 0$. 




Alternatively, the scaling of entanglement entropy with system size $L$ can be studied, as shown in Fig. \ref{fig:Fig2}(b) for the averaged half-chain entanglement entropy $\bar{S}^{(\gamma)}_L(L/2)$ (solid lines) and quarter-chain entanglement entropy $\bar{S}^{(\gamma)}_L(L/4)$ (dashed lines). When $\gamma$ is below the critical point (black, blue and orange line), a monotonous growth of entanglement entropy is observed when $L$ is increased and a significant difference can be distinguished between the curves of half-chain and quarter-chain entanglement entropy, relating back to the critical scaling of the lines seen in Fig. \ref{fig:Fig2}(a).
For larger $\gamma$, when dephasing dominates (green and red lines), both half-chain and quarter-chain entropy coincide and remain constant as a function of system size, thus reflecting the area law with a plateau of the residual entropy $s_0(\gamma)$ when $c(\gamma)\approx 0$ in \eqref{eq:S_cft}, shown in Fig. \ref{fig:Fig2}(a).

To study the behavior across the transition, we show in Fig. \ref{fig:Fig2}(c) the steady-state scaling of half-chain entropy $\bar{S}^{(\gamma)}_L(L/2)$ as function of $\gamma$ for different system sizes $L$. When $\gamma$ is below $\gamma_c$, the critical point we find in the Gutzwiller approach \cite{supp}, the curves for different $L$ fall apart. Upon increasing $\gamma$, $\bar{S}^{(\gamma)}_L(L/2)$ decreases for all $L$ and when a critical point is reached, close to $\gamma_c^\text{(GW)}=3$ from the Gutzwiller analysis \cite{supp}, the curves for different $L$ converge. For higher $\gamma$ the curves coincide, which confirms that $\bar{S}^{(\gamma)}_L(L/2)$ is uniform for different system sizes $L$ in the dephasing regime, shown in Fig. \ref{fig:Fig2}(b).

We believe that we have strong indications for critical scaling, in particular by satisfying \eqref{eq:c_alpha}. However, to unambiguously exclude the possibility of a volume law over critical scaling for $\gamma \rightarrow 0$, a thorough analysis of larger system sizes $L$ is required. Also topological entanglement entropy \cite{zabalo2020critical} could be a promising route. However, this quantity is prohibitively difficult to obtain with MPS simulations.

\begin{figure}
    \centering
    \includegraphics[width=\columnwidth]{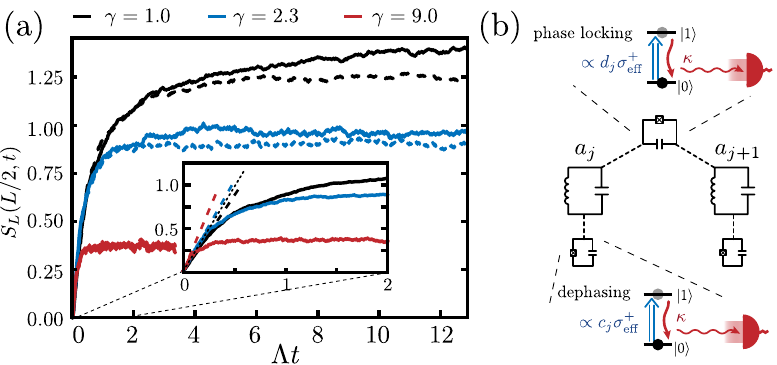}
    \caption{The time evolution of $\bar{S}^{(\gamma)}_{L}\big(\frac{L}{2},t\big)$ in time for $L=32$ (solid) and $L=16$ (dashed)  obtained from averaging over $500$ trajectories. Below the critical point (black and blue lines) entanglement entropy for different $L$ converges to different values, while above (red lines) it converges to the same steady value. In the inset we show the short-time behavior and see that the initial growth is linear (dashed lines), with a rate close to $\frac{\Lambda}{2}$ (dotted line). (b) A schematic of the setup proposed for the experimental implementation. The cavities are coupled 2-by-2 to disspative ancilla spins for the phase locking jumps and each cavity is coupled to another ancilla for the dephasing. Registering spontaneous spin decays in the ancillae allows for the registering of cavity jumps.}
    \label{fig:Fig3}
\end{figure}

Finally, Fig. \ref{fig:Fig3} shows the evolution of half-chain entanglement entropy $\bar{S}^{(\gamma)}_L\big(L/2,t\big)$ over time for $L=32$ (solid lines) and $L=16$ (dashed lines) for different values of $\gamma$. Starting from a zero entropy state, we let the system evolve and sample trajectories to see the rise in entanglement entropy. A saturation time $t_\text{st}$ is found where $\bar{S}^{(\gamma)}_L\big(L/2,t\big)$ converges to a steady-state value, schematically depicted in Fig. \ref{fig:Fig1}(b), which depends on both the system size $L$ and reduced dephasing rate $\gamma$. When dephasing is dominant (red lines),  $\bar{S}^{(\gamma)}_L\big(L/2,t\big)$ rapidly stabilizes and the curves for different $L$ are indistinguishable from each other, as expected for the area law. In the regime where phase-locking dominates (blue and black lines) the convergence is much slower, since entanglement spreads between distant sites. Different system sizes $L$ (solid vs. dotted lines) now converge to different steady-state values, reflecting the critical scaling of $\bar{S}^{(\gamma)}_L(l)$ \eqref{eq:S_cft}, previously shown more accurately in Fig. \ref{fig:Fig2}(b).

The initial growth, shown in the inset of Fig. \ref{fig:Fig3}, is close to linear (dashed lines), i.e. $\bar{S}^{(\gamma)}_L\big(L/2,t\big) = \kappa t$, with $\kappa \approx \frac{\Lambda}{2}$ (dotted line). The growth of trajectory entanglement entropy is thus reminiscent of the entanglement growth after a quench, where also an initial linear behavior is seen which saturates to a steady value. \cite{calabrese2005evolution}.

 \textit{Circuit QED implementation.--} Although originally presented in a cold-atom context \cite{diehl2008quantum}, phase-locking \eqref{eq:ch} can also be engineered in circuit QED \cite{marcos2012photon}. We propose the realization of a coupling between two adjacent cavities and an ancilla qubit  $H_\text{eff} \approx g_\text{eff} d_j \sigma_{j}^x$, with $\sigma^x = \sigma^+ + \sigma^-$. If the qubit is very lossy, registering a spontaneous qubit decay corresponds to detecting a phase-locking jump. In \cite{supp} we elaborate a scheme to engineer $H_\text{eff}$ by coupling the cavities 2-by-2 to a driven ancilla with an anharmonic level structure, such as a fluxonium qubit \cite{girvin2009circuit}.
 
While dephasing noise \eqref{eq:cd} is ubiquitous in quantum systems \cite{boissonneault2009dispersive,sears2012photon,schachenmayer2014spontaneous,yanay2014heating,bernier2020melting}, it is generally not possible to monitor the environment that induces the noise. In our approach, however, we keep track of individual trajectories, an essential aspect, and we propose a scheme to engineer dephasing processes by coupling each cavity to another lossy ancilla with $\sim a_j^\dagger a_j \sigma_x$. Upon registering an ancilla emission jump, one can infer the occurrence of a dephasing jump $c_j$, see \cite{supp}. This is in contrast with \cite{sun2014tracking}, where a coupling $H\sim a^\dagger a \sigma_z$ was used to monitor the cavity parity with qubit measurements to register photon decay. In our proposal, the ancilla serves both to engineer \emph{and} to register the dephasing jump $c_j$.

 In Fig. \ref{fig:Fig3}(b) we show a schematic of the proposal for the simultaneous realization of the two monitoring protocols by coupling two ancillae to each cavity, see \cite{supp}.
 
 \textit{Conclusions and outlook.--} We have investigated the scaling transition for entanglement entropy averaged over trajectory states $\bar{S}^{(\gamma)}\big(l)$ from two competing monitoring protocols. We report a transition in the steady-state trajectory entanglement entropy from area law to critical scaling \eqref{eq:S_cft}, where the central charge satisfies the relation for CFT's \ref{eq:c_alpha} for different R\'enyi entropies.
 
 
Investigating larger filling factor $n>1$ would allow for the study of entanglement entropy in different $U(1)$ charge sectors \cite{goldstein2018symmetry,murciano2020entanglement}. The unraveling of a master equation is not unique and, as such, entanglement depends on the monitoring \cite{nha2004entanglement}. It would be fascinating to investigate if a similar transition can be seen for different unravelings within the same master equation. Since trajectory entanglement entropy is a quantity that is challenging (if not impossible) to measure directly in experiment -- it requires identical copies of the same stochastic state \cite{elben2019statistical,pichler2016measurement} -- investigating if there could be a local probe to witness the transition, like in circuit models \cite{gullans2020scalable}, is an exciting question. Finally, it would be intriguing to see if quantum states can be stabilized with feedback from jumps in a continuous monitoring scheme \cite{wiseman1994quantum}. 
 
\begin{acknowledgments}
 We acknowledge stimulating discussions with Rosario Fazio during the initial formation of this project and subsequent fruitful insights by Oles Shtanko, Luis Pedro Garcia-Pintos, Alexey Gorshkov, Michael Gullans, Mohammad Maghrebi, Dolf Huybrechts and Michiel Wouters. MVR gratefully acknowledges support in the form of a BAEF postdoctoral fellowship. ZC, AS, HD, and MH were supported by AFOSR FA9550-19-1-0399, ARO W911NF2010232, W911NF-15-1-0397 and NSF Physics Frontier Center at the Joint Quantum Institute. We used the Bridges system, which is supported by NSF award number ACI-1445606, at the Pittsburgh Supercomputing Center (PSC) \cite{bridges1,bridges2}. The authors acknowledge the University of Maryland supercomputing resources (http://www.it.umd.edu/hpcc) made available in conducting the research reported in this paper.
 \end{acknowledgments}

 \bibliography{DPT}

\end{document}


\begin{center}
{\Huge\textbf  Supplemental Material}
\end{center}

\section{The circuit implementation}
In this section, we elaborate the experimental implementation of our model within an integrated superconducting circuit. Each cavity is weakly coupled to two ancilla systems, one associated with phase-locking and the other with dephasing monitoring (Eq. (1) and (2), respectively, from main text, illustrated in Fig. 3b). The central idea is that the corresponding jumps can be recorded as spontaneous emission events in the ancillary systems. 

We first elaborate in detail the realization of the phase-locking and dephasing jumps and then a simulation of the dephasing and phase-locking scheme is provided.

\subsection{Phase-locking}

\begin{figure}[h]
    \centering
    \includegraphics[width=.65\columnwidth]{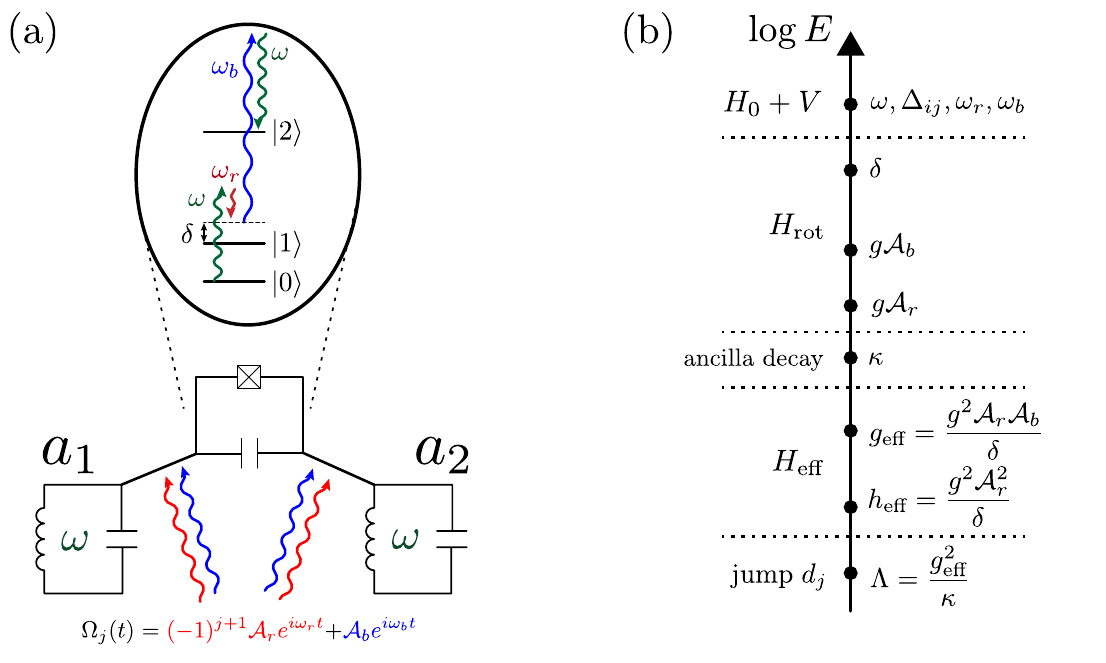}
    \caption{(a) A schematic of the experimental implementation of of the phase-locking jump between two cavities with modes $a_1$ and $a_2$. The full setup is irradiated with the two-tone beam $\Omega_j(t)$ to induce the transitions as given in the level diagram for the ancilla. (b) An overview of the hierarchy of energy scales necessary to get from $H_0 + V$, via a rotating wave approximation ($H_\text{rot}$), adiabatic elimination ($H_\text{eff}$) and a Born-Markov approximation, to the phase-locking jump operators $d_j$.}
    \label{fig:levels}
\end{figure}

We first illustrate the experimental scheme for monitoring the phase-locking jumps $d_j = (a^\dagger_j + a^\dagger_{j+1})(a_{j}-a_{j+1})$ for two cavities $a_1$ and $a_2$ coupled to an ancilla and will generalize the idea to a full chain at the end. A schematic illustration of the two-cavity setup can be found in Fig. \ref{fig:levels}(a). Recently, a closely related scheme was also elaborated for engineering two-photon dissipation to stabilize a photon pair condensate \cite{cian2019photon}.

Both cavities are coupled to a strongly anharmonic three-level system, which can be implemented on a circuit with fluxonium or transmon qubits \cite{girvin2009circuit}. The coupled Hamiltonian is of the form $H = H_0 + V$, with the free Hamiltonian
\begin{equation}
\label{eq:H0}
H_0 = \omega a^\dagger_1 a_1 + \omega a^\dagger_2 a_2 + \sum_{i = 0}^2 \epsilon_i |i\rangle\langle i |,
\end{equation}
where $\epsilon_i$ is the energy of the anharmonic oscillator at the $i$th level, and the coupling between the cavities and the anharmonic ancilla is induced by a CW drive
\begin{equation}
    V = \sum_{j = 1}^2 g(a_j +a_j^\dagger)(\Sigma + \Sigma^\dagger)(\Omega_j(t) + \Omega^*_j(t)),
    \label{eq:appendix_three_wave}
\end{equation}
where $\Sigma = |0\rangle \langle 1 | + \sqrt{2} |1\rangle \langle 2 |$ is the annihilation of the anharmonic oscillator and $\Omega_j(t)$, consisting of two tones, one is blue-detuned and the other red-detuned
\begin{equation}
\label{eq:drive}
\Omega_j(t) = (-1)^{j+1}\mathcal{A}_r e^{i\omega_r t} + \mathcal{A}_b e^{i\omega_b t},
\end{equation}
with $\omega_r = \omega - \Delta_{10}-\delta$ and $\omega_b =-\omega -\Delta_{21} +\delta $ and the ancilla energy level differences $\Delta_{ij} = \epsilon_i - \epsilon_j$. It is important that the amplitude $\mathcal{A}_r$ is antisymmetric, while $\mathcal{A}_b$ is symmetric.

Applying the rotating wave approximation in \eqref{eq:appendix_three_wave}, we find to leading order
\begin{equation}
\label{eq:H_rot}
H_\text{rot} = g\mathcal{A}_r(a_1-a_2)|1\rangle\langle 0 |e^{-i\delta t} + g\mathcal{A}_b(a^\dagger_1+a^\dagger_2)|2\rangle\langle 1 |e^{i\delta t} + h.c.
\end{equation}

When the detuning $\delta$ is much larger than $|g\mathcal{A}_r|$ and $|g\mathcal{A}_b|$ (weak-coupling regime), the state $|1\rangle$ can be adiabatically eliminated. For this we use Heisenberg equation of motion to find
\begin{eqnarray}
i\partial_t \big(|1 \rangle \langle 0| \big) &=& [|1 \rangle \langle 0|, H_\text{rot}] =i g e^{i\delta t} \bigg( A_b (a^\dagger_1 + a^\dagger_2 ) |2\rangle \langle 0 |  -  A_r (a^\dagger_1 - a^\dagger_2 ) \big(|1\rangle \langle 1|-|0\rangle \langle 0|\big) \bigg)\\
i\partial_t \big(|2 \rangle \langle 1| \big) &=& [|2 \rangle \langle 1|, H_\text{rot}] = ig e^{-i\delta t}\bigg( - A_r (a_1 - a_2 ) |2\rangle \langle 0 |+ A_r (a_1 + a_2 ) \big(|2\rangle \langle 2|-|1\rangle \langle 1|\big) \bigg).
\end{eqnarray}
If there is a strong decay $\kappa$ in the ancilla system (as we quantify below) it will almost always be found in the ground state $|0\rangle$ and the level occupations of $|1\rangle$ and $|2\rangle$ can be neglected, so that $\big\langle P_0 \big\rangle \approx 1$ and $\big\langle P_1 \rangle\approx \langle P_2 \rangle \approx 0$, with the projection operator on ancilla levels $P_n = |n\rangle \langle n|$.

The formal solutions are then found as
\begin{equation}
|1\rangle \langle 0| (t) = ig \int_0^t ds \bigg( A_b (a^\dagger_1 + a^\dagger_2 ) |2\rangle \langle 0 |  +  A_r (a^\dagger_1 - a^\dagger_2 ) \bigg) e^{i\delta s} , \;\;\; 
|2\rangle\langle 1| (t) = -ig A_r \int_0^t ds  (a_1 - a_2 ) |2\rangle \langle 0 | e^{-i\delta s} 
\end{equation}

Substituting these solutions in $H_\text{rot}$ \eqref{eq:H_rot} and assuming that $\delta$ is much larger than any frequency associated with the cavity-ancilla dynamics, we can make two approximations: (i) we send the integration boundary $t\rightarrow \infty$ (ii) we time-average over the fast oscillations with frequency $\delta$. 

After evaluation, we obtain a new effective Hamiltonian
\begin{equation}
\label{eq:H_eff_pl}
    H_\text{eff} \approx g_\text{eff} (a^\dagger_1+a^\dagger_2) (a_1-a_2) \sigma_\text{eff}^+ + h_\text{eff} (a_1-a_2)(a^\dagger_1-a^\dagger_2)|0\rangle\langle 0|  + h.c.,
\end{equation}
with the effective couplings $g_\text{eff} = \frac{g^2 \mathcal{A}_r \mathcal{A}_b}{\delta}$ and $h_\text{eff}=\frac{g^2 \mathcal{A}_r^2}{\delta}$, and $\sigma_\text{eff}^+\equiv |2\rangle \langle 0|$ the raising operator in the effective two level ancilla system. If we furthermore assume the decay rate $\kappa$ from $|2\rangle$ to $|0\rangle$ to be very large, i.e. $\kappa \gg g_\text{eff}$, the ancilla can be treated as a Markovian bath and the Born-Markov approximation can be applied to the first term $\propto g_\text{eff}$ \cite{gardiner2004quantum} to obtain the effective jump operators $d_1$ from the main text, with a decay rate $\Lambda = \frac{g_\text{eff}^2}{\kappa}$. The protocol of continuous monitoring now consists of detecting the photons from spontaneous emission of the transition $|2\rangle$ to $|0\rangle$ in the ancilla system. 

The second term in \eqref{eq:H_eff_dp}, scaling as $h_\text{eff}$, is the ac Stark shift and produces a level shift together with an effective hopping between the cavities. By choosing $A_b \gg A_r$, we can keep $g_\text{eff}$ (leading to the jumps) constant, while having only a small contribution from the ac Stark shift $h_\text{eff}$. Additionally, this spurious hopping can be canceled further by introducing an extra hopping barrier in \eqref{eq:H0}, i.e. $H_0 \rightarrow H_0 + J(a^\dagger_1 a_2 + a^\dagger_2 a_1)$ with matching $J=h_\text{eff}$, as can be engineered in circuit QED \cite{kounalakis2018tuneable}. 

This protocol can be extended straightforwardly to a whole chain of cavities, by coupling them 2-by-2 to ancilla three-level systems. The full system is then irradiated with the two-tone drive from \eqref{eq:drive}, where the $\mathcal{A}_r$ is of staggering order and $\mathcal{A}_b$ is uniform throughout the lattice. We this, the set of dissipators $(-1)^{j+1} d_j$ (1) from the main text will be found, with an irrelevant phase factor $\pm1$. Registering a photon click in the ancilla connecting site $j$ and $j+1$ then amounts to the detection of the jump $d_j$ in the cavity chain. In Fig. \ref{fig:levels}(b) we provide an overview of the energy scales introduced in this derivation to obtain the effective jump operator $d_1$.

\subsection{Controlled dephasing}

While dephasing is an omnipresent type of dissipation in many physical systems, for the purposes of this work, we want to engineer it in a controllable way throughout the chain to register the jumps $c_j = a^\dagger_j a_j$. Similar to \eqref{eq:H_eff_pl} for phase locking, here we need to engineer an effective coupling between each cavity and a new ancilla two-level system of the form 
\begin{equation}
\label{eq:H_eff_dp}
    H_\text{eff} = g_\text{eff} a^\dagger a \sigma_\text{eff}^x .
\end{equation}
In the strongly dispersive regime for the ancilla, where the Born-Markov approximation holds, the recording of a spontaneous spin decay event in the ancilla then corresponds to a dephasing jump $a^\dagger a$ in the cavity.

The same idea as for the phase-locking protocol could be followed, where each cavity is coupled to a strongly anharmonic three-level ancilla with a coupling of the form $V\sim (a+a^\dagger)(\Sigma_j+\Sigma_j^\dagger)(\Omega(t) + \Omega^\ast(t))$ and a drive $\Omega(t) = \mathcal{A}_r e^{i\omega_r t} + \mathcal{A}_b e^{i\omega_b t}$ that has the same relationship for the frequencies as \eqref{eq:drive}. The same analysis would then result in a coupling of the form $\sim a^\dagger a \sigma^x_\text{eff}$ for each cavity, leading to the dephasing jump operators of the form $c\sim a^\dagger a$ in the Born-Markov approximation and jumps that can be registered again as spontaneous emission events in the ancilla. The ac Stark shift would then simply correspond to a small (uniform) level shift of the cavity modes from $H_0$ \eqref{eq:H0}.

\subsection{Simulation of the monitoring scheme}
\begin{figure}
    \centering
    \includegraphics[width = .8\textwidth]{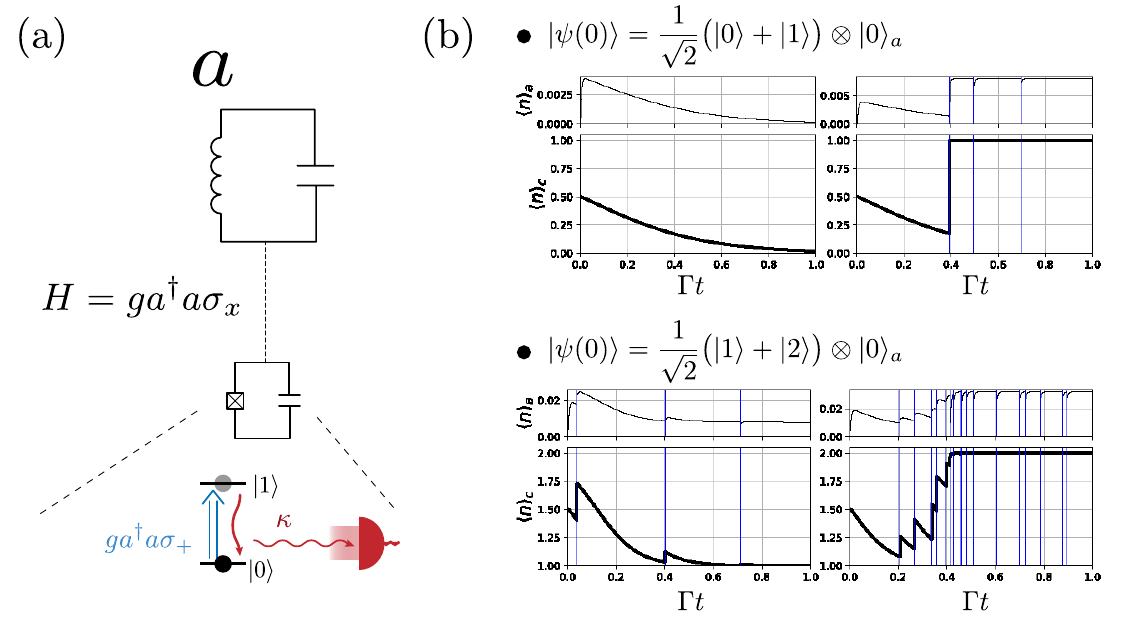}
    \caption{(a) Schematic of a cavity coupled to an ancilla qubit with coupling $H\sim a^\dagger a \sigma_x$ to record the dephasing jumps as ancilla decay jumps. (b) Lower panels show cavity occupation $\langle n \rangle_c$ starting from two different initial states that are a superposition of two number states, with the ancilla in state $|0\rangle$. Upper panels show the ancilla qubit occupation $\langle n \rangle_a$. Blue vertical lines are ancilla decay jumps. For each case, two trajectories are plotted that converge to lower number state (left panels) or the higher number state (right panel) of the initial superposition. The first case (left) shows few decay jumps, while the second (right) shows many jumps. }
    \label{fig:deph}
\end{figure}

We start with providing a simulation of the scheme for dephasing, for which a cavity is coupled to an ancilla qubit with a coupling $H=g a^\dagger a \sigma_x$, as illustrated in \ref{fig:deph}(a). The ancilla has a large decay rate $\kappa = 500 g^2/\Gamma$, with $\Gamma$ the effective dephasing rate. In Fig \ref{fig:deph}(b) we show trajectories for an initial state that is a product state of a superposition of two number states for the cavity and the ground state for the ancilla, i.e. $|\psi(0)\rangle = \frac{1}{\sqrt{2}}\big(|n_1\rangle + |n_2\rangle\big)\otimes |0\rangle_a$, with $n_1<n_2$. The stochastic evolution for the cavity number $\langle n \rangle_c$ (lower panels) and ancilla number $\langle n \rangle_a$ (upper panels) are shown and the ancilla decay jumps are indicated with blue vertical lines. Dephasing jumps $a^\dagger a$ favor higher number states, while the anti-Hermitian $H = -\Gamma/2 a^\dagger a$ decreases the norm of these, bringing the cavity to lower number states, so that the overall probability to end in state $|n_1\rangle$ or $|n_2\rangle$ for the initial state is $1/2$. 

When monitoring the ancilla decay jumps, we see this as well; many jumps (blue vertical lines) in the ancilla lead to $|n_2\rangle$, while few jumps (and therefore anti-Hermitian evolution dominates) lead to $|n_1\rangle$. If we start from a superposition of $|0\rangle$ and $|1\rangle$, only one jump is needed to bring you to $|1\rangle$, as we also see in the picture after a single ancilla decay (upper right). We thus conclude that dephasing jumps on the cavity can be inferred by monitoring the spontaneous emission events of the ancilla qubit. Note also that $\langle n \rangle_a \approx 0$ at all times (upper panels) due to the large ancilla decay rate, which allows us to make the Born-Markov approximation to effectively obtain the dephasing jumps.

We would like to stress here once more that the standard monitoring scheme from quantum non-demolition (QND) measurements relies on a coupling of the form $H\sim a^\dagger a \sigma_z$, which is inherently different from our proposal. In the QND case, the evolution of the cavity number can be inferred by measuring the qubit resonance repeatedly in time, since the number of particles in the cavity results in a level shift for the qubit. In the same spirit, the parity of a cavity was monitored by performing qubit measurements at a high repetition rate after applying a controlled phase gate \cite{sun2014tracking}. However, in this work the cavity state is always projected upon a parity subspace, such that cavity losses can be inferred, whereas our primary interest lies in tracking the occurrence of the same quantum jumps that we engineer with the ancilla coupling, \emph{without} performing a (partially) projective measurement.

\begin{figure}
    \centering
    \includegraphics[width=.7\columnwidth]{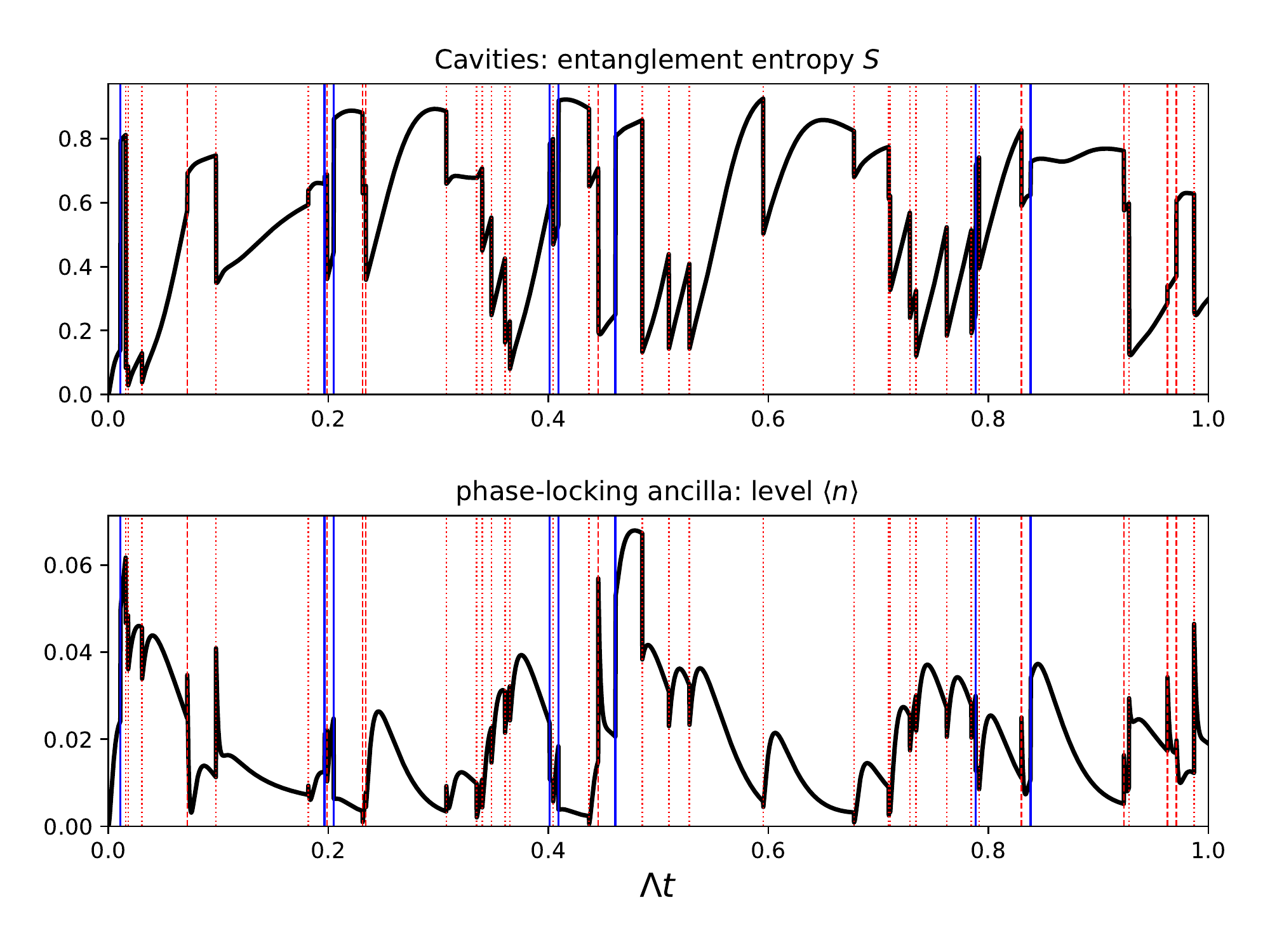}
    \caption{An example of a trajectory for two cavities coupled to the phase-locking ancilla with local dephasing. The upper panel shows the entanglement entropy $S$ of one cavity with the other, while the lower panel shows the ancilla expectation $\langle n \rangle$. Decay jumps in the phase-locking ancilla are indicated with blue lines, while red lines correspond to dephasing jumps in the cavities. }
    \label{fig:phl_traj}
\end{figure}

In Fig. \ref{fig:phl_traj} we show a simulation of the setup depicted in Fig. 3b of the main article. We consider two cavities that are coupled to an ancilla qubit with $H_\text{eff}$ \eqref{eq:H_eff_pl}, where we neglect ac Stark shift ($g_\text{eff}\gg h_\text{eff}$). The ancilla qubit is very lossy, with a decay rate $\kappa = 500 g^2_\text{eff}/\Lambda$. In addition, each cavity is subject to local dephasing $c_j = a^\dagger_j a_j$ with a rate $\Gamma = 10\Lambda$.

The upper panel from Fig. \ref{fig:phl_traj} shows the Von Neumann entropy of the first cavity $S$, quantifying its amount of entanglement with the second (or the other way around), while the lower panel shows the ancilla qubit occupation $\langle n \rangle$. Since the decay of the qubit $\kappa$ is very large, it is almost always found in the ground state $|0\rangle$. When an ancilla decay jump is detected (blue lines) we see an abrupt rise in $S$ (upper panel) and, of course, a sudden drop of the ancilla to $|0\rangle$ -- this corresponds to applying the phase-locking jump $d_j$ to the cavities in the Born-Markov approximation \cite{gardiner2004quantum}. On the other hand, if a dephasing jump (red lines) is recorded in cavity $1$ (dotted) or cavity $2$ (dashed) we (almost always) see an abrupt drop of $S$ and a rise of ancilla $\langle n \rangle$. Therefore, this illustrates once more that dephasing and phase-locking are incompatible and that a phase-locking jump might evoke a dephasing jump and vice versa, resulting often in a series of jumps within a short time, as can be seen in the trajectory around $\Lambda t \approx [0.2, 0.4, 0.8]$. Also interesting, when no jump is detected over a period of time, we still see continuous dynamics in both $S$ and $\langle n \rangle$, as would be found from the evolution with the anti-Hermitian cavity Hamiltonian from the quantum jump approach. As a consequence, both the detection \emph{and} and the absence of a quantum jump yield information about the quantum state for an observer that tracks the dynamics of the system with a continuous monitoring scheme. In conclusion, the detection of a decay jump in the ancilla coupled to the two cavities corresponds to the recording of a phase-locking jump $d_j$.

\section{The Gutzwiller master equation}
\label{sec:gutz}
For the Gutzwiller approach we start from the reduced density matrix of a trajectory state $|\psi\rangle$, defined as $\rho_j\big[|\psi \rangle \big] = \text{tr}_{\bar{j}} \big( |\psi\rangle \langle \psi|\big)$, where $\text{tr}_{\bar{j}}\cdot$ means tracing over all sites other than $j$. After averaging over trajectory states $|\psi \rangle$, we find an effective master equation for the averaged $\rho_j$ only, which is of the form
\begin{equation}
\label{eq:rho_j}
\partial_t \rho_j = \text{tr}_{\bar{j}} \big\{\partial_t \rho \big\} = \text{tr}_{\bar{j}} \big\{\mathcal{L}(\rho)\big\},
\end{equation}
where $\rho$ is the full system density matrix and $\mathcal{L}(\rho)$ the full Liouvillian of the system, 
\begin{equation}
\mathcal{L}(\rho) = \sum_i \gamma_i \Big( c_i \rho c^\dagger_i- \frac{1}{2}(c^\dagger_i c_i\rho + \rho c^\dagger_i c_i) \Big),
\end{equation}
with $c_i$ the jump operators and $\gamma_i$ the corresponding dissipation rates rates.

While it is easy to check that the master equation is invariant under a transformation to new jump operators $\tilde{c}_i$ of the form $\tilde{c}_i \equiv \sum_j U_{ij} c_j$ with $U$ some unitary matrix, we emphasize that the validity of the Gutzwiller approach to the master equation is intimately related to the unraveling associated with our monitoring scheme, with the \emph{local} jumps (1) and (2) from the main text. In other words, it is exactly this what motivates us to interpret $\rho_j$ as the averaged on-site reduced density matrix from the trajectories arising in our unraveling.

\subsection{Phase locking}

The phase-locking jumps $d_{j}$ act only on sites $j$ and $j+1$. The Gutzwiller approach now relies on the assumption that one can approximate the two-site reduced density matrix as $\rho_{j,j+1}=\rho_j\otimes\rho_{j+1}$. This leads to the effective on-site Liouvillian for the jump operator $d_{j}$ of the local density matrix $\rho_j$
\begin{equation}
\label{eq:onsite}
\tilde{\mathcal{L}}^\text{(pl)}_j(\rho_j) =  \text{tr}_{j+1} \bigg\{ d_{j} (\rho_j\otimes \rho_{j+1}) d^\dagger_{j} -\frac{1}{2} \Big( d^\dagger_{j} d_{j} (\rho_j\otimes \rho_{j+1}) +  (\rho_j\otimes \rho_{j+1}) d^\dagger_{j} d_{j} \Big)\bigg\} 
\end{equation}

We then find a mean-field master equation for the local density matrix $\rho_j$ (we omit the index `$j$' and the factor `$2$' comes from the two jumps $d_{j-1}$ and $d_{j}$ acting on site $j$)
\begin{equation}
\label{eq:MEpl}
\partial_t \rho_j\Big|_\text{pl} = 2\tilde{\mathcal{L}}^\text{(pl)}(\rho_j).
\end{equation}
After working out \eqref{eq:onsite}, we find that
\begin{equation}
\tilde{\mathcal{L}}^\text{(pl)}(\rho) =  \tilde{\mathcal{L}}^\text{(pl,d)}(\rho) + \tilde{\mathcal{L}}^\text{(pl,e)}(\rho) + \big(\tilde{\mathcal{L}}^\text{(pl,e)}(\rho)\big)^\dagger,
\end{equation}
with
\begin{equation}
\mathcal{L}^\text{(pl,d)}(\rho) = n \,\Big( a^\dagger \rho a - \frac{1}{2}\big\{ a a^\dagger,\rho\big\}\Big) 
+(n+1) \Big( a \rho a^\dagger - \frac{1}{2}\big\{ a^\dagger a,\rho\big\} \Big) 
 + a^\dagger a \rho a^\dagger a - \frac{1}{2}\big\{ a^\dagger a a^\dagger a, \rho \big\} ,
\end{equation}
where $n = \langle a^\dagger a \rangle = 1$, in our case, is the mean on-site particle number.

We furthermore find the contribution for generating non-diagonal order in $\rho_j$
\begin{eqnarray}
\nonumber
\mathcal{L}^\text{(pl,e)}(\rho) &=& \frac{1}{2}\big( \langle a^\dagger a a^\dagger\rangle + \langle a^\dagger a^\dagger a\rangle \big) \big( \rho a - a\rho \big) - \langle a^2\rangle \Big( a^\dagger \rho a^\dagger - \frac{1}{2} \big\{ a^\dagger a^\dagger, \rho \big\} \Big)\\
\label{eq:Le}
&& + \langle a \rangle\Big( a^\dagger a \rho a^\dagger - \frac{1}{2}\big\{ a^\dagger a^\dagger a, \rho \big\} -a^\dagger \rho a^\dagger a + \frac{1}{2} \big\{ a^\dagger a a^\dagger, \rho \big\} \Big).
\end{eqnarray}

It is now easy to check that a coherent state $|\alpha\rangle$, having $a|\alpha\rangle = \alpha |\alpha\rangle$, is a dark state of the local master equation -- it annihilates the r.h.s. of \eqref{eq:MEpl}. We can find the norm of $\alpha$ from the filling factor as $|\alpha|=\sqrt{n}=1$, while the phase is free and set by the initial state.

\subsection{Dephasing}

The dephasing is much more straightforward because it is local and acts on single sites only. For the local density matrix we then find 
\begin{equation}
\partial_t \rho \Big|_\text{dp} = \mathcal{L}^{(dp)}(\rho) =  a^\dagger a \rho a^\dagger a - \frac{1}{2}\big\{ a^\dagger a a^\dagger a, \rho \big\}
\end{equation}

\subsection{Full time evolution and the steady state}
\begin{figure}
    \centering
    \includegraphics[width=.7\columnwidth]{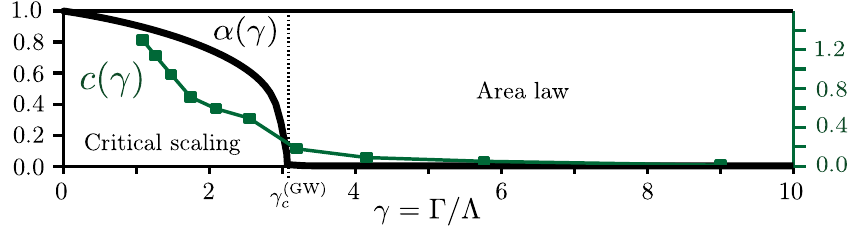}
    \caption{A comparison between the Gutzwiller order parameter $\alpha$ (black solid) and the effective central charges $c(\gamma)$ from the main text (green squares). For the order parameter $\alpha$ we see a sharp transition across a critical point, where it vanishes. At this point also a strong reduction is seen of $c(\gamma)$, marking the critical scaling vs. area law transition for the trajectory entanglement entropy. }
    \label{fig:compare_GW}
\end{figure}

The full time-evolution with the correct monitoring rates of the on-site density matrix $\rho$ is then found as
\begin{equation}
\label{eq:full_GW}
\partial_t \rho = 2\eta\tilde{\mathcal{L}}^\text{(pl)}(\rho) + \gamma \tilde{\mathcal{L}}^\text{(dp)}(\rho),
\end{equation}
which we can numerically integrate in time with a Runge-Kutta scheme to obtain expectation values of local observables $\langle O \rangle_t = \text{tr}\big[\rho(t) O]$.

Alternatively, one can look at the equation of motion for local operator expectation values directly, by evaluating $\partial_t\langle O \rangle = \text{tr}\big[\partial_t\rho O\big] = \text{tr}\big[\mathcal{L}(\rho) O\big]$, giving the time dependence of the order parameter
\begin{equation}
\label{eq:GW_a}
\partial_t \alpha = 2 \Lambda \big( \langle a^\dagger a^2 \rangle - \langle a^2 \rangle \alpha^\ast \big) - \frac{\Gamma}{2} \alpha.
\end{equation}

Here we can readily identify the dark states in two limiting cases. First, when $\Gamma=0$, there is only phase locking and the steady state is given by $a|\alpha \rangle = \alpha |\alpha \rangle$, which is the case for a coherent state for all the sites. Second, if $\Lambda=0$, the pure dephasing dynamics lead to $\alpha \rightarrow 0 $ in the long-time limit, i.e., non-diagonal order disappears.

We can now perform a full numerical analysis of the Gutzwiller master equation \eqref{eq:full_GW} and evaluate the steady-state value of the order parameter $\alpha$ after long enough integration times, as shown in Fig. \ref{fig:compare_GW}. The order parameter (black solid line) shows a sharp transition across a threshold value $\gamma \approx 3$, where it vanishes. In the limit $\gamma_c^{(GW)} \rightarrow 0$ we obtain $\alpha = \sqrt{n} = 1$, the known result for a local coherent state. When we compare with the effective central charges $c(\gamma)$ (green squares) obtained from the entropy profiles, we see that they are strongly suppressed at the point $\gamma_c^{(GW)}$ where the $\alpha$ vanishes. 

Though, we stress that we interpret this more as a suggestive qualitative indication, rather than a rigorous quantitative analysis, given that we are numerically unable to perform a proper scaling analysis for the finite-system entanglement entropy profiles. Moreover, we suspect that taking the thermodynamic limit in 1D for the trajectory states, as is presumed for the Gutzwiller analysis, would lead to a buildup of quantum fluctuations that destroy the order parameter and create quasi-long-range order, rather than full condensation -- reminiscent of what is seen in equilibrium models \cite{kuhner1998phases}.

%